\documentclass[11pt]{article}
\usepackage{braket}
\usepackage{graphicx}
\usepackage[fleqn]{amsmath}
\usepackage{color}

\usepackage[top=2.5cm,bottom=2.5cm,left=2.5cm,right=2.5cm]{geometry}
\usepackage{cite}
\begin{document}

\title{An Open-Source Framework for $N$-Electron Dynamics: II. Hybrid Density Functional Theory/Configuration Interaction Methodology}

\date{\today}

\author{Gunter Hermann \thanks{corresponding author} \thanks{Institut f\"ur Chemie 
und Biochemie, Freie Universit\"at Berlin, Takustra{\ss}e 3, 14195 Berlin, Germany} 
\thanks{These authors contributed equally to this work.},
Vincent Pohl\footnotemark[2] \footnotemark[3], 
and Jean Christophe Tremblay\footnotemark[2]}

\maketitle

\section{Abstract}

In this contribution, we extend our framework for analyzing and visualizing 
correlated many-electron dynamics to non-variational, highly scalable electronic structure method.
Specifically, an explicitly time-dependent electronic wave packet is written as a linear combination of 
$N$-electron wave functions at the configuration interaction singles (CIS) level,
which are obtained from a reference time-dependent density functional theory (TDDFT) calculation.
The procedure is implemented in the open-source Python program \textsc{detCI@ORBKIT},
which extends the capabilities of our recently published post-processing toolbox [{\it J.\,Comput.\,Chem.}~{\bf37}\,(2016)\,1511].
From the output of standard quantum chemistry packages using atom-centered Gaussian-type basis functions, 
the framework exploits the multi-determinental structure of the hybrid TDDFT/CIS wave packet 
to compute fundamental one-electron quantities such as 
difference electronic densities, transient electronic flux densities, and transition dipole moments.
The hybrid scheme is benchmarked against wave function data for the laser-driven state selective excitation in LiH.
It is shown that all features of the electron dynamics are in good quantitative agreement with the higher-level method
provided a judicious choice of functional is made.
Broadband excitation of a medium-sized organic chromophore further demonstrates the scalability of the method.
In addition, the time-dependent flux densities unravel the mechanistic details of
the simulated charge migration process at a glance.

\clearpage

\section{Introduction}

Unraveling the flow of electrons inside a molecule out of equilibrium is key to understand its reactivity.
Since the pioneering laser experiments by Zewail and co-workers\cite{87:zewail,88:zewail}, the development of new light sources 
has now granted access to the indirect observation of electron dynamics on its natural timescale. To shed light
on the mechanistic details of this attosecond dynamics, accurate theoretical methods are required that capture
the subtle details of the transient electronic structure evolution. Various approaches based on explicitly 
time-dependent density functional theory (TDDFT) and wave function ansatz have been developed over the years and enjoyed mixed degrees of success.
While TDDFT appears as more intuitive and scalable, it was shown to suffer from problems for ultrafast dynamics in 
strong laser fields.
On the other hand, the advantages of wave function-based methods in terms of convergence
become rapidly compensated by their unfavorable computational cost.
Further, the intuitive picture of electrons flowing on a molecular skeleton can become blurred by correlation effects
between the $N$ particles.

This contribution is motivated by the need for a robust, scalable wave function method to investigate ultrafast $N$-electron
dynamics in systems of large dimension. The method we advocate is based on a combination of linear-response time-dependent functional theory
(LR-TDDFT) and configuration interaction singles (CIS) methodologies, as was introduced recently \cite{11:SCS:tddftci,hermann2016alianatase,klinkusch2016roi}.
In principle, the method is similar to the well-established CIS ansatz, with the exception that the energies and the pseudo-CIS
eigenvectors are obtained from a reference LR-TDDFT calculation. This allows to improve the energetic properties of the states
while keeping a simple electron-particle picture to describe the transient $N$-electron wave packet. This TDDFT/CIS hybrid formalism
inherits the qualities of both underlying methods and ensures the $N$-representability of all reduced density matrices,
at all times and under all laser conditions. The ensuing $N$-electron dynamics remains marred by the non-intuitive interpretation
of quantities beyond the density itself.

Recently, we demonstrated that correlated electron dynamics can be accurately described by means of the electronic flux density
operator and derived one-electron properties.
We introduced an open-source framework \cite{Pirhadi2016127} to post-process multi-determinantal configuration interaction wave functions directly from the output of
standard quantum chemistry packages.
It thus becomes possible to reconstruct the transient $N$-representative one-electron density and current density (flux density) using a library of transition moments
calculated from the multi-determinantal configuration interaction wave functions, yielding an intuitive tool for visualizing and analyzing the correlated electron dynamics.
A wide variety of established wave function-based methods are covered, ranging from configuration interaction singles to Full CI via 
restricted active space CI and multi-configuration self-consistent-field methods.
It is the purpose of this work to extend the formalism to the TDDFT/CIS hybrid formalism mentioned above, 
which should retain the qualities of the wave function ansatz and the scalability of DFT-based schemes with respect to the system size.

In the next section, the hybrid TDDFT/CIS methodology is first introduced, followed by the description of the analysis toolset
based on the flux density. The application section reports on benchmark calculations on the LiH molecule and the demonstration
of the scalability of the scheme by investigation of the broadband excitation in an organic chromophore. The findings are summarized 
in the conclusion section.
Unless otherwise stated, atomic units are used throughout the manuscript ($\hbar = m_e = e = 4\pi\varepsilon_0 = 1$).

\section{Theory}\label{theory}
\subsection{Hybrid TDDFT/CI Methodology}

The evolution of the electronic state of a molecular system obeys the time-dependent Schr{\"o}dinger equation\cite{Schroedinger1926quantisierung},
which can be written in the clamped nuclei approximation
\begin{eqnarray}\label{TDSE}
i\frac{\partial}{\partial t} \Ket{\Psi_{{\rm el}}\left( t\right)} = \left(\hat{H} - \hat{\mu}\cdot\vec{F}(t)\right) \Ket{\Psi_{{\rm el}}\left( t\right)}.
\end{eqnarray}
The interaction of the molecular dipole $\hat{\mu}$ with an external laser field $F(t)$ is treated here semi-classically.
For a system consisting of $N$ electrons and $N_A$ nuclei, the field-free electronic Hamiltonian reads
\begin{eqnarray}\label{hamiltonian}
    \hat{H} = - \frac{1}{2} \sum\limits_{i=1}^{N} \nabla_i^2 + \sum\limits_{i=1}^{N} \sum\limits_{j>i}^{N} \frac{1}{r_{ij}} - \sum\limits_{i=1}^{N} 
\sum\limits_{A=1}^{N_A} \frac{Z_A}{r_{Ai}}\text{,}
\end{eqnarray}
where $\frac{1}{r_{ij}}=\frac{1}{|\vec{r}_i - \vec{r}_j|}$ is inter-electronic Coulomb repulsion, and $r_{Ai}$ is the distance between the $i$th electron and nucleus $A$ of charge $Z_A$.
In this work, an electronic wave packet $\Ket{\Psi_{{\rm el}}\left( t\right)}$  satisfying Eq.~\eqref{TDSE}
is expressed as a linear superposition of stationary electronic states $\Ket{\Phi_\lambda}$
\begin{eqnarray}\label{td_wf}
\Ket{\Psi_{{\rm el}}\left(t\right)} = \sum_{\lambda} B_{\lambda}\left( t\right) \Ket{\Phi_\lambda} .
\end{eqnarray}
Here, $B_{\lambda}\left( t\right)$ are the expansion coefficients of state $\lambda$, which describe the time-evolution of the wave packet.
For molecules in strong laser fields, a large number of stationary electronic states is required to offer a proper description of the $N$-electron dynamics.
The equations of motion for the coefficients in Eq.~\eqref{TDSE}, associated with the basis set expansion Eq.~\eqref{td_wf}, can be integrated numerically.

In the time-dependent configuration interaction methodology, the stationary electronic states are chosen as 
linear combinations of excited configuration state functions
\begin{eqnarray}\label{general_CI}
\Ket{\Phi_{\lambda}^{{\rm CI}}} = C_{\textrm{ref}}^{\left(\lambda\right)} \Ket{\phi_{\textrm{ref}}}+\sum_{ar}C_{a}^{r\left(\lambda\right)}\Ket{\phi_{a}^{r}}
+\sum_{abrs}C_{ab}^{rs\left(\lambda\right)}\Ket{\phi_{ab}^{rs}} + \ldots\,.
\end{eqnarray}
The expansion parameters $C^{\left(\lambda\right)}$ are associated with the formal excitation 
of a reference configuration, $\Ket{\phi_{\textrm{ref}}}$, from occupied orbitals $\left\lbrace a,b,c \right\rbrace$ 
to virtual orbitals $\left\lbrace r,s,t \right\rbrace$.
Including all possible excitations leads to the exact Full CI limit.
The reference and excited configurations are defined as Slater determinants,
which builds antisymmetrized products of one-electron spin orbitals $\Ket{\varphi_a}$.
Note that, in the time-dependent configuration interaction (TDCI) methodology in the form presented above, the field-free electronic Hamiltonian
is considered to be diagonal in the basis of CI eigenstates at a given level of theory. The matrix elements of the dipole operator
can be computed from the knowledge of these eigenfunctions, which serve as a basis for the variational representation of the molecule-field interaction.

For large molecules, it is customary to truncate the CI expansion to a chosen maximum rank of excitations (e.g., CI Singles or CI Singles Doubles)
in order to reduce the number of possible excited configurations. Unfortunately, this often compromises the energetic description of the excited 
states. To circumvent this limitation while keeping the problem computationally tractable, Sonk and Schlegel \cite{11:SCS:tddftci} first recognized that  
only excitation energies and transition dipole moments are required to perform TDCI simulations. These can be obtained from linear-response 
time-dependent density functional theory (LR-TDDFT).
To generalize this approach, it was proposed to use the solutions of the LR-TDDFT calculation
to generate a basis of pseudo-CI eigenstates \cite{klinkusch2016roi,hermann2016alianatase}. 
All required information for a TDCI simulation is thus available 
from the output of standard quantum chemistry programs, provided excitations are performed from the ground state.

According to the Runge-Gross theorem, it is possible to recast the $N$-electron Schr{\"o}dinger equation and calculate
all observables from the sole knowledge of the one-electron density. Using the Kohn-Sham ansatz for the density,
the $N$-electron time-dependent  Schr{\"o}dinger equation can be mapped onto a one-electron equation for the orbitals
\begin{eqnarray}
i\frac{\partial\varphi_{a}(\mathbf{r},t)}{dt} = \left(-\frac{\nabla^2}{2} + v_{\textrm{KS}}(\mathbf{r},t)\right)\varphi_{a}(\mathbf{r},t).
\end{eqnarray}
The time-dependent Kohn-Sham potential $v_{\textrm{KS}}(\mathbf{r,t})$ contains the classical electrostatic interaction ($v_{\textrm{Hartree}}(\mathbf{r},t)$),
an external potential ($v_{\textrm{ext}}(\mathbf{r},t)$), and an exchange-correlation contribution ($v_{\textrm{xc}}(\mathbf{r},t)$), i.e.,
\begin{eqnarray}\label{Vks}
v_{\textrm{KS}}(\mathbf{r,t})=v_{\textrm{Hartree}}(\mathbf{r},t)+v_{\textrm{ext}}(\mathbf{r},t)+v_{\textrm{xc}}(\mathbf{r},t).
\end{eqnarray}
In explicit TDDFT, the Kohn-Sham potential is usually assumed to be local in time.
A celebrated success of TDDFT comes from its linear-response formulation, which allows to accurately
compute spectral properties of large molecules. For this endeavor, the response kernel of the electron density
to an external weak, long wavelength perturbation can be evaluated from the electric susceptibility of the ground state.
The search for the poles of the response function can be recast as an eigenvalue problem of the form
\begin{eqnarray}\label{casida}
\left[\left(\begin{array}{rr} \mathbf{A}&\mathbf{B}\\ \mathbf{B}^{\dagger}&\mathbf{A}^{\dagger}\end{array}\right)-\omega\left(\begin{array}{rr} -\mathbf{I}&0\\ 0&\mathbf{I}\end{array}\right)\right] \left(\begin{array}{c} \mathbf{X}\\\mathbf{Y}\end{array}\right) 
=  -\left(\begin{array}{c} \delta \mathbf{v}\\ \delta \mathbf{v}^{\dagger}\end{array}\right),  
\end{eqnarray}
where $\delta \mathbf{v}$ is the response of the system state to the perturbation, $\delta v_{\textrm{ext}}(\mathbf{r},t)$.
The elements of matrices $\mathbf{A}$ and $\mathbf{B}$ are obtained from the orbital energies and integrals over the 
exchange-correlation kernel, see Eq.~\eqref{Vks}. At the resonance frequencies $\omega$, where the response vanishes ($\delta \mathbf{v}=0$),
the solution of the Casida Eq.~\eqref{casida} yields simultaneously the excitations and 
de-excitations amplitudes, $\mathbf{X}$ and $\mathbf{Y}$. 
In the present work, we make use of the fact that these are usually given in the output of standard quantum chemistry
programs, together with the excitation energies and the oscillator strengths.

From Eq.~\eqref{casida}, it is possible to define pseudo-CI Singles eigenvectors in the Tamm-Dancoff approximation, which consists 
in neglecting the off-diagonal blocks $\mathbf{B}$. This procedure can alter the quality of the energetic properties of the excited states.
On the other hand, the dominant characters present in the pseudo-CI eigenvectors are often not strongly affected by this approximation.
In the TDDFT/CI procedure, we thus advocate using directly the transition energies and amplitudes obtained from a LR-TDDFT calculation
to take advantage of the full solution of Eq.~\eqref{casida} and to obtain a good energetic description of the excited states.
A separate Tamm-Dancoff calculation may be used to confirm the character of the excited states.
The LR-TDDFT excitation amplitudes are then re-orthonormalized using a modified Gram-Schmidt procedure
to define a pseudo-CI basis for the TDCIS dynamics.
All properties not directly deriving from the energies can be subsequently calculated at the CIS level of theory using the 
orbitals and the pseudo-CI eigenstates, which are treated as configuration interaction singles expansions.
Note that the Slater determinants $\left\lbrace \phi_{\textrm{ref}}, \phi_{a}^{r} \right\rbrace$  
are constructed from Kohn-Sham orbitals.
Importantly, all the information required to reconstruct these KS-orbitals and the $N$-electron pseudo-eigenfunctions
are directly accessible from the output of standard quantum chemistry packages.
As a consequence, only the evaluation of one-electron integrals is required to generate a library of molecular properties and 
transition moments of various one-electron operators, which can be used to characterize the properties of transient wave packets,
as explained below.

\subsection{Analysis Tools for Electron Dynamics}

For the analysis of the $N$-electron dynamics, we propose using a set of tools composed from the one-electron density,
$\rho \left(\mathbf{r},t \right)$, and the associated electronic flux density, $\mathbf{j} \left(\mathbf{r},t\right)$.
These are related by the electronic continuity equation
\begin{eqnarray}\label{con_eq}
\frac{\partial}{\partial t} \rho \left(\mathbf{r},t \right) = - \vec{\nabla} \cdot \mathbf{j} \left(\mathbf{r},t\right).
\end{eqnarray}
Whereas the electron density gives information about the probability distribution of the electron,
the flux density yields complementary information about the phase of the electronic wave packet.
This in turn reveals the mechanistic aspects of the time-evolution of the one-electron density.
The one-electron density can be used to define the electron flow, $\frac{\partial}{\partial t} \rho \left(\mathbf{r},t \right)$,
as the left-hand-side of the continuity equation. The difference density, $\mathbf{y} \left(\mathbf{r},t \right)$, is a widespread
quantity used for visualization purposes, and it can be obtained by integrating the electron flow from a chosen initial condition
$\rho \left(\mathbf{r},0 \right)$, i.e.,
\begin{eqnarray}\label{diff_dens}
  \mathbf{y} \left(\mathbf{r},t \right) &= \int_{0}^{t} {\rm d} t' \frac{\partial \rho \left(\mathbf{r},t' \right)}{\partial t'} = \rho \left(\mathbf{r},t \right) - \rho \left(\mathbf{r},0 \right).
\end{eqnarray}
We will resort to both quantities in later analyses.

In operator form, the one-electron density and the electronic flux density respectively read
\begin{eqnarray}
\hat{\rho} \left( \mathbf{r}\right) &= \sum_{k}^{N} \delta\left( \mathbf{r} - \mathbf{r}_{k} \right)= \sum_{k}^{N} \delta_k(\mathbf{r})\label{rho_op}\\
\hat{j} \left( \mathbf{r}\right) &= \frac{1}{2}\sum_{k}^{N} \left(\delta_k(\mathbf{r})\hat{p}_{k} + \hat{p}^{\dagger}_{k}\delta_k(\mathbf{r})\right)\label{j_op},
\end{eqnarray}
where $\mathbf{r}$ is an observation point, $\delta\left( \mathbf{r} - \mathbf{r}_{k} \right)= \delta_k(\mathbf{r})$ is the Dirac delta distribution
at the position $\mathbf{r}_{k}$ of electron $k$, and $\hat{p}_{k}=-i\vec{\nabla}_{k}$ is the associated momentum operator.
In general, the expectation value of any one-electron operator $\hat{F}$  can be expressed using Eqs.~\eqref{td_wf} and \eqref{general_CI} as 
\begin{eqnarray}
\Braket{\hat{F}}(t) & = & \Braket{\Psi_{{\rm el}}\left(t\right) \left| \hat{F} \right| \Psi_{{\rm el}}\left(t\right)} \label{exp_val_1} \\
& = & \sum_{\lambda\nu} B_{\lambda}^{\dagger}\left( t\right) B_{\nu}\left( t\right) \Braket{\Phi_{\lambda} \left| \hat{F} \right| \Phi_{\nu}} \label{exp_val_2}.
\end{eqnarray}
Evaluation of the matrix elements $\Braket{\Phi_{\lambda} \left| \hat{F} \right| \Phi_{\nu}}$ can be done by exploiting the structure 
of the functions $\left\{\Phi_{\lambda},\Phi_{\nu}\right\}$. In the hybrid TDDFT/CIS methodology, these take the form of singly excited 
configurations, i.e., the truncation of Eq.~\eqref{general_CI} at the singles level. The matrix elements in the basis 
of singly excited configurations read
\begin{eqnarray}
\Braket{\Phi_{\lambda} \left| \hat{F} \right| \Phi_{\nu}} & = & 
  C_{\textrm{ref}}^{\left(\lambda\right)}C_{\textrm{ref}}^{\left(\nu\right)}  \Braket{\phi_{\textrm{ref}}\left| \hat{F} \right| \phi_{\textrm{ref}} }
+ \sum_{bs}C_{\textrm{ref}}^{\left(\lambda\right)} C_{b}^{s\left(\nu\right)} \Braket{\phi_{\textrm{ref}} \left| \hat{F} \right| \phi_{b}^{s}}\nonumber\\
&& + \sum_{ar}C_{a}^{r\left(\lambda\right)}C_{\textrm{ref}}^{\left(\nu\right)}  \Braket{\phi_{a}^{r}\left| \hat{F} \right| \phi_{\textrm{ref}} }
+ \sum_{abrs}C_{a}^{r\left(\lambda\right)} C_{b}^{s\left(\nu\right)} \Braket{\phi_{a}^{r} \left| \hat{F} \right| \phi_{b}^{s}} \label{exp_val_3}.\\
 & = & C_{\textrm{ref}}^{\left(\lambda\right)}C_{\textrm{ref}}^{\left(\nu\right)} \sum_{a}\Braket{\varphi_{a} \left| \hat{F} \right| \varphi_{a}}
 + \sum_{ar}C_{a}^{r\left(\lambda\right)}C_{a}^{r\left(\nu\right)}  \sum\limits_{\overline{a}}\Braket{\varphi_{\overline{a}}\left| \hat{F} \right| \varphi_{\overline{a}} } \nonumber\\
&& + \sum_{bs}C_{\textrm{ref}}^{\left(\lambda\right)} C_{b}^{s\left(\nu\right)} \Braket{\varphi_{b} \left| \hat{F} \right| \varphi_{s}}
 + \sum_{ar}C_{a}^{r\left(\lambda\right)}C_{\textrm{ref}}^{\left(\nu\right)}  \Braket{\varphi_{a}\left| \hat{F} \right| \varphi_{r} } \nonumber\\
&&+ \sum_{ar\ne s}C_{a}^{r\left(\lambda\right)} C_{a}^{s\left(\nu\right)} \Braket{\varphi_{r} \left| \hat{F} \right| \varphi_{s}}
+ \sum_{a\ne br}C_{a}^{r\left(\lambda\right)} C_{b}^{r\left(\nu\right)} \Braket{\varphi_{a} \left| \hat{F} \right| \varphi_{b}} \label{exp_val_4}.
\end{eqnarray}
where $\overline{a} \in \{1,2,\dots,a-1,r,a+1,\dots\}$ denote the occupied spin orbitals of the configuration state function $\Ket{\phi_{a}^{r}}$.
Note that we make use of the Slater-Condon rules\cite{slater1929theory,condon1930theory,slater1931theory} to resolve Eq.~\eqref{exp_val_4}
in terms of one-electron integrals in the basis of the spin orbitals, $\varphi_{a}(\mathbf{r})$. 
The transition moments between spin orbitals are usually computed in the spin-free representation by first integrating over the spin coordinates.
Specifically, the expectation value for the electron density requires the following integrals
\begin{eqnarray}\label{rho_mo}
\Braket{\varphi_{a} \left| \hat{\rho} \right| \varphi_{b}} = \rho_{ab}(\mathbf{r})=\varphi_{a}(\mathbf{r})\varphi_{b}(\mathbf{r}).
\end{eqnarray}
The electronic flux density for a wave packet of the form Eq.~\eqref{td_wf} can be formulated as
\begin{eqnarray}\label{j}
\mathbf{j} \left(\mathbf{r},t\right) & = & 2 i\sum_{\lambda<\nu}
{\rm Im}\left[B^{\dagger}_{\lambda}\left(t\right)B_{\nu}\left(t\right)\right] \mathbf{J}_{\lambda \nu} \left(\mathbf{r},t\right) \label{j_matrix} ,
\end{eqnarray}
which can be calculated by exploiting the CIS structure of the eigenfunctions.
The transition electronic flux density from state $\lambda$ to state $\nu$ is denoted $\mathbf{J}_{\lambda \nu} \left(\mathbf{r},t\right)$,
which simplifies using the Slater-Condon rules to 
\begin{eqnarray}\label{j_CI_mo}
\mathbf{J}_{\lambda \nu} \left(\mathbf{r},t\right) &=& 
\sum_{ar} \Big(C_{\textrm{ref}}^{\left(\lambda\right)} C_{a}^{r\left(\nu\right)} + C_{a}^{r\left(\lambda\right)}C_{\textrm{ref}}^{\left(\nu\right)}\Big)  \mathbf{j}_{ar} \nonumber\\
&&+ \sum_{a,r\ne s}C_{a}^{r\left(\lambda\right)} C_{a}^{s\left(\nu\right)} \mathbf{j}_{rs}
 + \sum_{a\ne br}C_{a}^{r\left(\lambda\right)} C_{b}^{r\left(\nu\right)} \mathbf{j}_{ab}\label{transitionJ}
\end{eqnarray}
where 
\begin{eqnarray}\label{j_mo}
\mathbf{j}_{ab} = -\frac{i}{2}\Big(\varphi_{a}(\mathbf{r})\vec{\nabla}\varphi_{b}(\mathbf{r}) - \varphi_{b}(\mathbf{r})\vec{\nabla}\varphi_{a}(\mathbf{r})\Big)
\end{eqnarray}
are molecular orbital (MO) electronic transition flux densities from MO $\varphi_{a}(\mathbf{r})$ to MO $\varphi_{b}(\mathbf{r})$.

As one of the most widespread bases used in quantum chemistry, we specialize here to spatial MO defined 
as linear combination of atom-centered  orbitals (MO-LCAO for ``Molecular Orbital - Linear Combination of Atomic Orbitals'')
\begin{eqnarray}\label{molcao}
\varphi_{a} \left(\mathbf{r}\right) = \sum_{A=1}^{N_{A}}\sum_{i_A=1}^{n_{{\rm AO}}{(A)}} D^{(a)}_{i_A} \chi_{i_A}\left(\mathbf{r}-\mathbf{R}_{A}\right),
\end{eqnarray}
where $D^{(a)}_{i_A}$ is the $i_A$th expansion coefficient for MO $a$.
The atomic orbitals $\chi_{i_A}$ are expressed as a function of the Cartesian coordinates of one electron $\mathbf{r}$
and the spatial coordinates $\mathbf{R}_{A}$ of nucleus $A$.
$N_{{A}}$ labels the number of atoms and $n_{{\rm AO}}{(A)}$ is the number of atomic orbitals on atom $A$.
Using the MO-LCAO ansatz, the transition moments between spin orbitals read
\begin{eqnarray}
\Braket{\varphi_{a} \left| \hat{F} \right| \varphi_{b}} =
 \sum_{A,B}^{N_{{A}}}\sum_{i_A=1}^{n_{{\rm AO}}{(A)}}\sum_{j_B=1}^{n_{{\rm AO}}{(B)}} D^{(a)}_{i_A} D^{(b)}_{j_B} \Braket{\chi_{i_A} \left| \hat{F} \right| \chi_{j_B}}.
\end{eqnarray}
The MO-LCAO coefficients $D^{(a)}_{iA}$ and the definition of the atomic orbitals can be read directly from the output of standard quantum chemistry program packages.
All required derivatives and integrals in the atomic orbital basis are computed analytically using our Python post-processing toolbox \textsc{ORBKIT}\cite{orbkit},
with which the molecular orbital density (cf. Eq. \eqref{rho_mo}) and the molecular orbital electronic flux density (cf. Eq. \eqref{j_mo}) can then be projected on an arbitrary grid.
Combining the information in this list with the occupation patterns of the quasi-CI eigenvectors associated with the excited states obtained at the LR-TDDFT level
of theory, it is possible to create a library of transition moments between CI-states to be used in the dynamics. 
Note that the transition dipole moments are also computed using the same information and exploiting the multi-determinantal 
structure of the $N$-electron basis functions, cf. Eqs.~\eqref{td_wf} and \eqref{general_CI}.
The analysis tools for the hybrid TDDFT/CIS methodology are implemented, along with various other one-electron quantities,
in a recently introduced open-source Python framework \textsc{detCI@ORBKIT}, available at \texttt{https://github.com/orbkit/orbkit/}.
The program requires a preliminary LR-TDDFT calculation using 
Gaussian-type atom-centered orbitals. There is no restriction for the choice of functional.
Currently, the code supports the GAMESS\cite{Gamess} and TURBOMOLE\cite{TURBOMOLE} formats.
Our program then computes matrix elements of one-electron operators, projects them on an arbitrary grid, and stores them in a library
to be used for analyzing the $N$-electron dynamics. 
The framework \textsc{detCI@ORBKIT} is written in Python, simplifying its portability on different platforms and offering efficient standard libraries
for visualization purposes. Implementation details are given elsewhere \cite{ci_orbkit_wf}. 
The dynamics program is not part of the standard implementation and can be performed using either a user-written code or, e.g., the Matlab
\textsc{WAVEPACKET} package\cite{wavepacket}. 
In the present work, all dynamical simulations were performed using \textsc{GLOCT}, an in-house implementation
of a propagator for the reduced-density matrix and related quantities \cite{tremblay2008time,08:TS:gloct}.

\section{Results and Discussion}

To demonstrate the capabilities of \textsc{detCI@ORBKIT}, we perform the analysis of correlated electron dynamics
in two selected molecular systems. First, the charge transfer process in lithium hydride is studied to benchmark
the quality of the TDDFT/CIS description against Full CI results.
The electron migration in an alizarin dye induced by broadband laser excitation is then used as an example to
demonstrate the scalability of the method.

\subsection{Benchmark: Charge Transfer in LiH}\label{lih}

\begin{figure}
\includegraphics[width=0.6\textwidth]{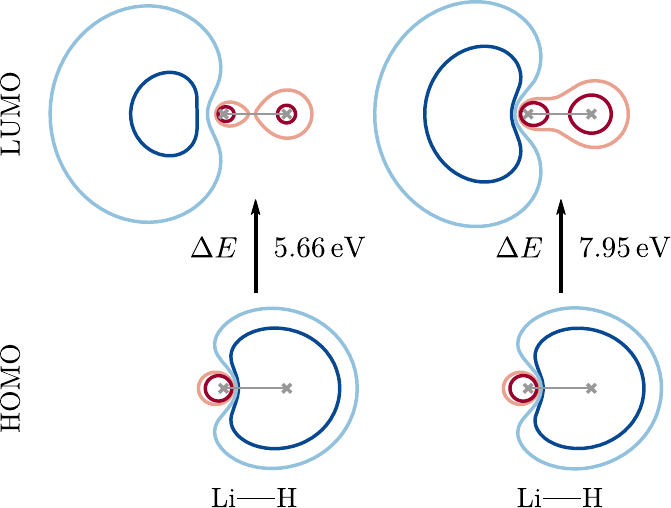}
\caption{Two-dimensional projections of the dominant orbitals involved in the excitation of the first charge transfer 
state in LiH. The HOMO is found to be similar at Hartree-Fock (left) and at the B2-PLYP (right) levels of theory. 
The LUMO obtained from B2-PLYP is more delocalized and found at significantly higher energy than the comprable
Hartree-Fock orbital.
}
\label{lih_mos}
\end{figure}

In lithium hydride, charge migration can be initiated, e.g.,
by laser excitation from the molecular ground state $\Ket{\Psi_{\rm g}}$ to the first excited state $\Ket{\Psi_{\rm e}}$.
The charge transfer mechanism can be understood from the analysis of the superposition state
\begin{eqnarray}
\Ket{\Psi_{{\rm el}}\left(t\right)} = \frac{1}{\sqrt{2}} \left(\Ket{\Psi_{\rm g}} e^{-i  E_{\rm g} t/\hbar} +\Ket{\Psi_{\rm e}} e^{-i (E_{\rm e} t/\hbar + \eta)}\right)
\end{eqnarray}
which leads to the time-dependent one-electron density
\begin{eqnarray}\begin{aligned}\label{rho_lih}
  \rho\left(\mathbf{r},t\right) =& \frac{1}{2} \Big(\rho_{\rm g}\left(\mathbf{r}\right) + \rho_{\rm e}\left(\mathbf{r}\right)\Big)
  + \rho_{\rm ge}\left(\mathbf{r}\right)\cos(\Delta E\,t/\hbar + \eta),
\end{aligned}\end{eqnarray}
where $\Delta E=E_{{\rm e}} -E_{{\rm g}}$, and $\eta$ is the relative phase. It is set to $\eta=\pi$ in the present example.
Similarly, the electronic flux density takes the form
\begin{eqnarray}\begin{aligned}\label{j_lih}
  \mathbf{j} \left(\mathbf{r},t\right) =  \mbox{Im}\left[\mathbf{J}_{\rm ge} \left(\mathbf{r}\right)\right] \sin(\Delta E\, t/\hbar + \eta),
\end{aligned}\end{eqnarray}
where the transition electronic flux density $\mathbf{J}_{\rm ge} \left(\mathbf{r}\right)$ is obtained from Eq.~\eqref{j_CI_mo}.
The charge transfer dynamics can be thus rationalized in terms of the static transition moment between the two states involved.
The electron densities, $\rho_{\rm g}\left(\mathbf{r}\right)$ and $\rho_{\rm e}\left(\mathbf{r}\right)$, of the ground state X${}^1 \Sigma^+$
and the charge transfer state A${}^1 \Sigma^+$ and the transition density between
both, $\rho_{\rm ge}\left(\mathbf{r}\right)$, are computed by combining the MO contributions obtained from Eq.~\eqref{rho_mo}. 
The Kohn-Sham orbitals, the pseudo-CI eigenvectors, and the associated LR-TDDFT excitation energies are computed
using an aug-cc-pVTZ at the B2-PLYP level of theory, as implemented in TURBOMOLE.\cite{TURBOMOLE}
The character of the charge transfer state is found to be dominated by the HOMO-LUMO transition (see Fig.~\ref{lih_mos} (right side)).
This is in good agreement with the character determined from CIS and Full CI calculations, 
both performed with PSI4\cite{psi4} using the identical basis set.
The corresponding frontier orbitals from the Hartree-Fock reference are shown in the left side of Fig.~\ref{lih_mos}.
It can be seen that the HOMO is similar in both cases, while the LUMO is more delocalized at the B2-PLYP level of theory.
We will show below that this difference has only a marginal influence on the electron dynamics.

\begin{figure}
\includegraphics[width=0.9\textwidth]{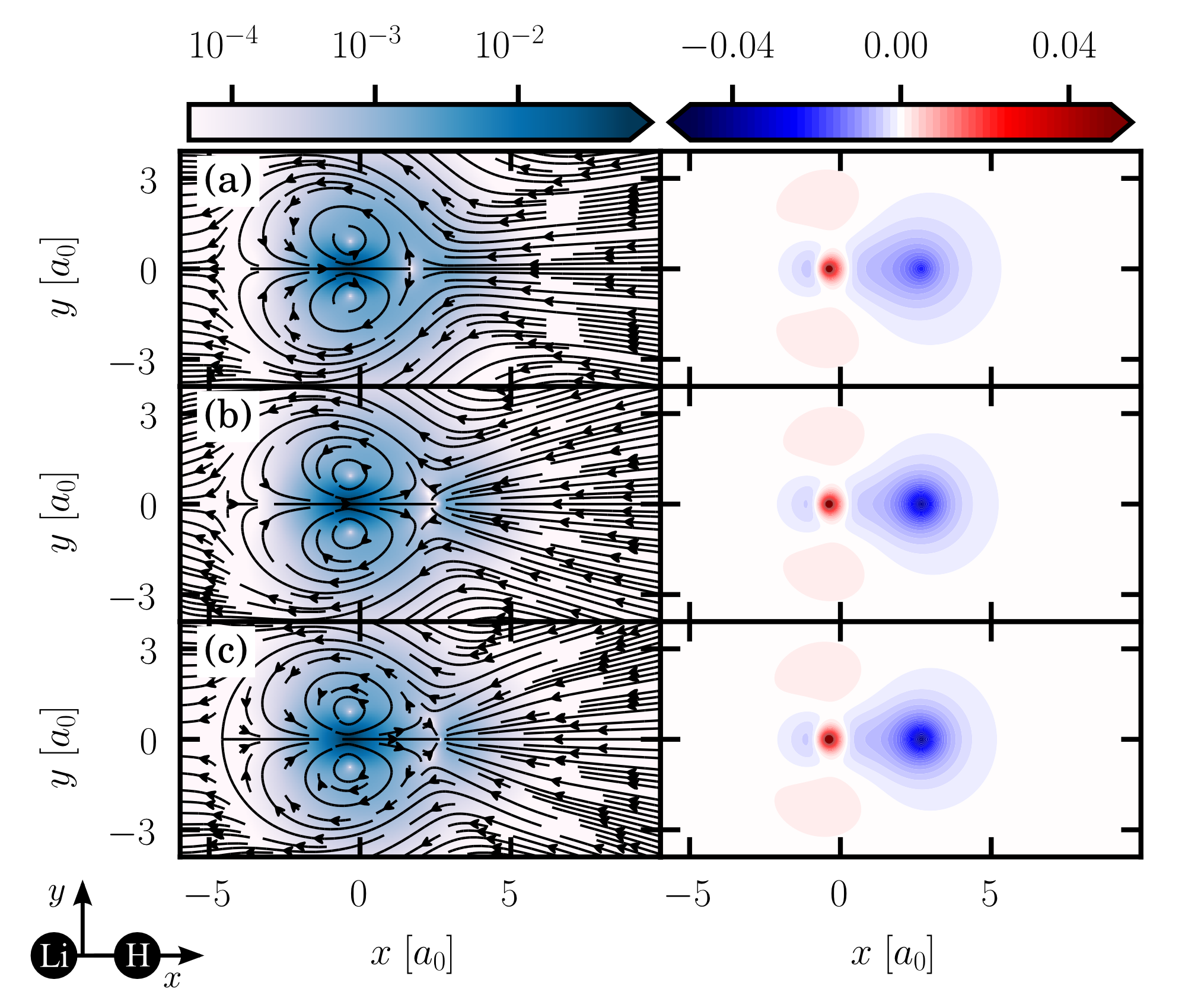}
\caption{Comparison of the electron transfer process in LiH at time $t=\tau/4$, 
obtained using different electronic structure methods
(a) Full CI; (b) CI Singles; (c) hybrid TDDFT/CIS using the B2-PLYP functional.
Left panels: the black arrows show the electronic flux density, $\mathbf{j} \left(\mathbf{r},t\right)$,
represented as streamlines. The color bar indicates the regions of high flux density (dark blue) and
low flux density (white) in units of $E_{\rm h}/\hbar a_0^2$. 
Right panels: the difference density in units of $a_0^{-3}$, Eq.~\eqref{diff_dens}, illustrates that an electron is transferred from
the hydrogen (density depletion in blue) to the lithium ion (density increase in red).
All qualitative features are in good agreement for all methods.
}
\label{lih_flux_dens}
\end{figure}

A great advantage of LR-TDDFT over CIS is the improved energetic description of the excited states at virtually the same computational cost.
This is demonstrated by the good agreement of the excitation energies at the B2-PLYP level of theory ($\Delta E=3.51$\,eV) with the Full CI
reference ($\Delta E=3.56$\,eV), as compared to the rather poor value for the truncated CIS expansion ($\Delta E=4.04$\,eV).
Since the excitation is dominated by single excitations in all cases, the transition energies will affect mostly the timescale of the dynamics.
Further, considering the similarities between the MOs involved in the dominant transition, a similar dynamical behavior is to be expected.
This is indeed what is observed in Fig.~\ref{lih_flux_dens}, where the left panels report the flux density $\mathbf{j} \left(\mathbf{r},t\right)$ 
at time $t=\tau/4$ computed using Eq.~\eqref{j_lih} and the right panels show the difference density $\mathbf{y}\left(\mathbf{r},t\right)$ computed from the one-electron density, Eq.~\eqref{rho_lih}.
In the top right panel, the benchmark Full CI calculation reveals that the charge is transferred from the hydrogen atom (electron density depletion region in blue)
to the lithium ion (red region). Some $p$-like regions of increasing density can also be recognized around the lithium atom. 
The same features are also observed for both the CIS (central panels) and hybrid (lower panels) method,
while the magnitude of the difference density is larger around the hydrogen for the two single electron approaches.

The electronic flux density maps are depicted as streamlines in the left panels of Fig.~\ref{lih_flux_dens}, where the blue shades indicate its magnitude.
The charge is seen to flow from the polarizable hydrogen around and towards the back of the hard lithium. The same large vortex around the lithium ion is observed for 
the Full CI benchmark, the CIS, and the hybrid TDDFT/CIS approach. 
The main quantitative differences between the methods are located in the low-density regions, e.g., at $x>5~a_0$, where the Full CI benchmark predicts 
a flux almost parallel to the molecular axis. The critical point (at $x\sim 2~a_0$) between the lithium and hydrogen atoms also appears to be slightly shifted to the right
at the CIS and TDDFT/CIS level of theory.
In general, it can be said that both single electron excitation ansatz yield a very similar picture of the electron difference density and the associated flux density. 
This conclusion is likely to hold for all dynamical processes involving $N$-electron eigenstates dominated by a single excitation character.

\subsection{Scalability: Electron Migration in Alizarin}\label{ali}

In this second example, we demonstrate the computational scalability of the hybrid TDDFT/CIS approach to analyze 
the correlated electron dynamics for more extended molecular systems.
The necessity of such a method is due to the fact that high-level electronic structure 
theory methods, such as MCSCF, are often not applicable for larger molecules.
The CI scheme truncated at the singles excitation represents a simple, intuitive, 
and computationally cheap approach to compute qualitatively correct excited electronic states.\cite{foresman1992toward,dreuw2003long} 
However, it yields inaccurate vertical excitation energies from the ground state.\cite{dreuw2005single}
As advocated in the theory section above, a suitable alternative is LR-TDDFT\cite{gross1990time}, which usually provides better energetic description than CIS while 
retaining the same quality for the wave function.
It can be inferred from the example in the previous subsection, that this will provide an adequate description for a large number of photochemical processes
dominated by a single excitation character.
In addition, it benefits from the versatility and continuous improvement of density functionals.
Proper treatment of the excited states strongly depends on the appropriate choice of a
functional, which can be chosen to correctly describe electronic correlations, the dispersive nature, or the charge-transfer character of a given excitation.
\cite{casida1998molecular,iikura2001long,heyd2003hybrid,tawada2004long,rohrdanz2008simultaneous,rohrdanz2009long,jacquemin2009extensive}
Fortunately, extensive experience has been accumulated over the years concerning the applicatibility of each functional in specific chemical contexts.
For example, it is known that non-local exchange improves greatly the description of charge transfer states \cite{dreuw2003long}.

\begin{figure*}
(a)\includegraphics[width=0.45\textwidth]{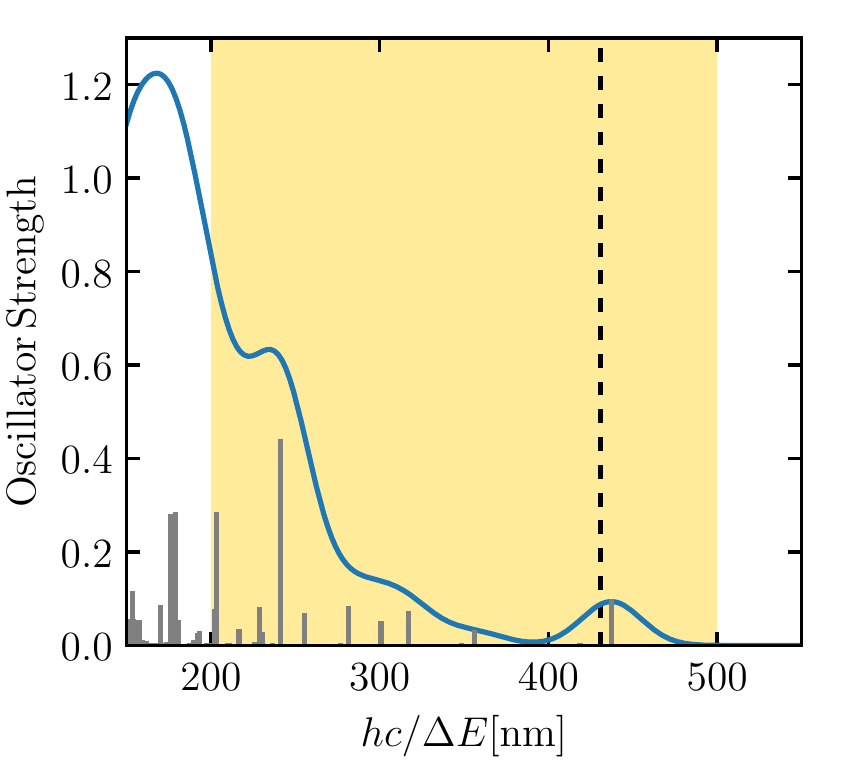}
(b)\includegraphics[width=0.45\textwidth]{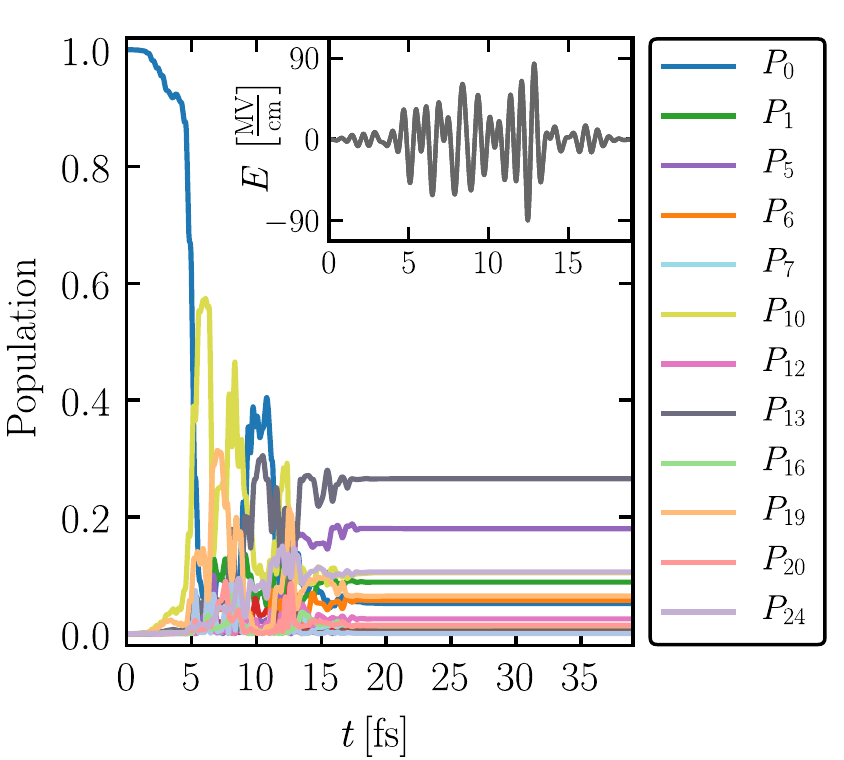}
\caption{
(a) Simulated optical spectrum of alizarin obtained from linear-response TDDFT
using the B3LYP functional and a def2-SVP basis set. 
The vertical gray lines represent the oscillator strengths of specific transitions. 
The broadened spectrum (solid blue line) is constructed using Gaussian functions with width $\sigma= 25\,{\rm nm}$.
The position of the experimental band of the free alizarin at 431 nm is marked by a dashed black line.
(b) Population evolution of the ultrafast $N$-electron dynamics in alizarin driven by broadband laser excitation. 
The inset shows the time evolution of the laser pulse. All states between 200\,nm and 500\,nm are excited from the ground state.
}
\label{ali_spectra_pop}
\end{figure*}

To show the scalability of the hybrid TDDFT/CIS formalism for the analysis of correlated 
electron dynamics, we initiate a photoinduced ultrafast charge migration process in alizarin.
This organic chromophore is used as a $\pi$-conjugated photosensitizer in dye-sensitized solar cells.
Prior to the dynamical simulation, the electronic and optical properties of alizarin
are determined by means of LR-TDDFT.
Therefore, we perform a TURBOMOLE\cite{TURBOMOLE} calculation with the B3LYP\cite{becke1988density}
hybrid functional and the def2-SVP basis set\cite{schafer1992fully,weigend2005balanced} 
at the equilibrium geometry of alizarin.
This setup has been previously proven to yield accurate results for the electronic 
and spectroscopic properties of such systems.\cite{duncan2005electronic,sanchez2010real,Gomez_2015,hermann2016alianatase}
The computed absorption spectrum is depicted in Fig. \ref{ali_spectra_pop}(a) along with 
the experimentally observed absorption band of free alizarin (dashed black line).
The good agreement for the first absorption band between theory (437 nm) and experiment 
(431 nm) underlines the suitability of TDDFT to model the electronic spectra of medium-sized organic 
molecules dominated by single excitation character. It is important to recognize that the second absorption band
is composed of a multitude of excited states in the UV/VIS range.

In order to simulate an ultrafast charge migration process in alizarin, we proceed to
the broadband excitation of all excited states in the energy range between 200\,nm and 500\,nm 
(cf. Fig.\,\ref{ali_spectra_pop}(a) yellow filling).
For the promotion of these states from the ground state, a superposition of state-to-state
sin$^2$-shaped pulses with a duration 19 fs is constructed.
The pulse is adjusted to the parameters of a realistic experimental laser field used
in similar investigations to initiate, e.g., ultrafast photoinduced processes in alizarin-TiO$_2$ solar cells.\cite{huber2002real,dworak2009ultrafast}
The resulting electric field is shown in the inset of Fig.\,\ref{ali_spectra_pop}(b).
The laser excitation is followed by a 20 fs field-free propagation.
The time-evolution of the $N$-electron wave packet (cf. Eqs.~\eqref{TDSE} and \eqref{td_wf})
is accomplished using an adaptive Runge-Kutta algorithm in the interaction picture.
The methodology and implementation details are described elsewhere.\cite{tremblay2004using,tremblay2008time,tremblay2011dissipative}
Fig. \ref{ali_spectra_pop}(b) shows the evolution of the state populations and the applied laser field in the inset.
As it is often the case for molecules in strong fields, the population dynamics is very intricate while the laser is on,
in part due to important polarization effects and in part due to the number of states that are excited by the broadband laser.
To account for the electronic response of the system to the laser field, 25 eigenstates are incorporated in the simulation.
After the laser excitation, only twelve states are significantly populated ($P_{\lambda}>0.01$).

To unravel the mechanistic pathways and give an intuitive picture of the electron dynamics, we advocate using
the time-dependent electron density, electron flux density, and electronic flow, which are reconstructed from the $N$-electron wave packet
in the pseudo-CI eigenvector basis.
This can become computationally tedious, since the number of Slater determinants 
in the wave function expansion increases with the number of occupied and unoccupied orbitals.
For alizarin in the current basis set, the $62$ occupied MOs times $248$ virtual MOs correspond to $15\,376$ determinants for each 
of the excited states. Recalling that reconstruction of the flux density, Eq.~\eqref{j_CI_mo}, requires 
combining one-electron integrals of all orbitals pairwise, for each pair of eigenstates,
this amounts to a tremendous computational task.
However, inclusion of the complete set of determinants is not necessary, since the 
expansion coefficient of many of these is either zero or negligibly small.
Moreover, the pseudo-CI eigenvectors are usually dominated by a few determinant contributions.
Applying the maximum correspondence principle between determinants and exploiting the Slater-Condon rules
lead to significant numerical savings, which are automatized in our implementation and can be controlled
by a user-defined convergence threshold.

In order to examine the influence of truncating the complete set of determinants to a 
physically meaningful subset in the wavefunction ansatz, we choose to use a threshold for the expansion coefficients
of the quasi-CI wave functions, Eq.~\eqref{general_CI}. All determinants with a contribution under this value are neglected
in the evaluation of the one-electron density and the associated flux density.
We define three different thresholds, $\left|C_a^{r,(\lambda)}\right| > \{10^{-3},\,10^{-2},\,10^{-1}\}$.
To illustrate the computational savings that can be expected,
the thirteenth excited state is selected as an example, 
since it is the most populated state after the broadband laser excitation.
The three chosen threshold values retain 926, 31, and 7 Slater determinants in the wave function expansion, respectively.
For consistency check, the pruned expansions recover 99.94 $\%$, 99.31 $\%$, and 95.92 $\%$ of the norm of the thirteenth excited state, respectively.
These numbers are similar for other excited states.
After excluding the contributions lying under the respective threshold, the remaining coefficients are renormalized using a modified Gram-Schmidt algorithm.
Regarding the computational effort, the reduction of determinants means a drastic decrease of computational steps.
For example, the calculation of the transition electronic flux density $\mathbf{j}_{5\,13} \left(\mathbf{r}\right)$
between the fifth and thirteenth excited state requires $N_{\rm SD}^{5} \times N_{\rm SD}^{13} =  573\times 926 = 530\,598$ 
determinant combinations for the more strict threshold of $\left|C_a^{r,\lambda}\right| >10^{-3}$.
This number reduces to $N_{\rm SD}^{5} \times N_{\rm SD}^{13} = 3\times7=21$ determinant combinations for the lenient threshold $\left|C_a^{r,\lambda}\right| >10^{-1}$.
This simple constraint thus confers great scalability to the method presented here.

\begin{figure*}
(a)\includegraphics[width=0.45\textwidth]{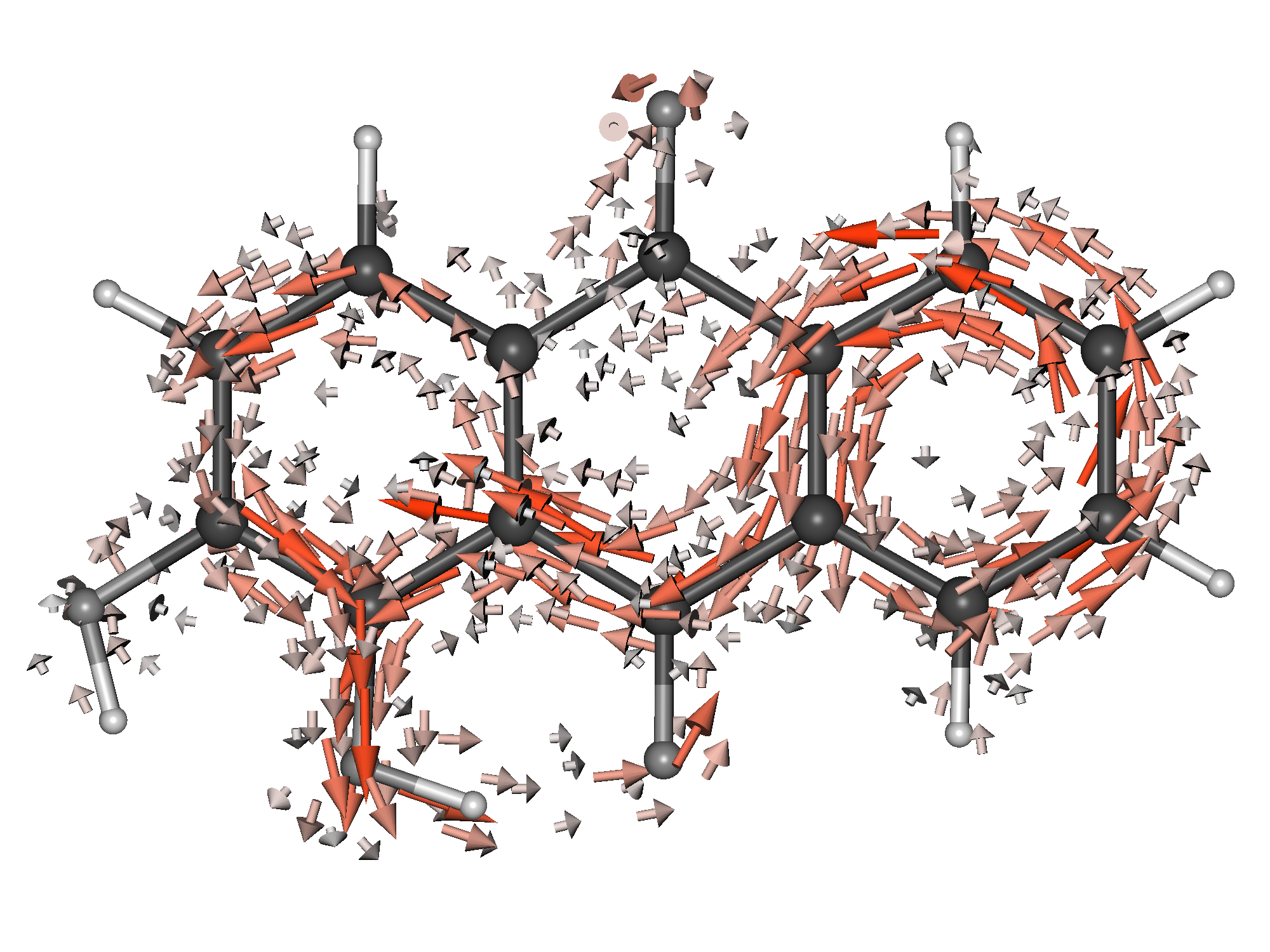}
(b)\includegraphics[width=0.45\textwidth]{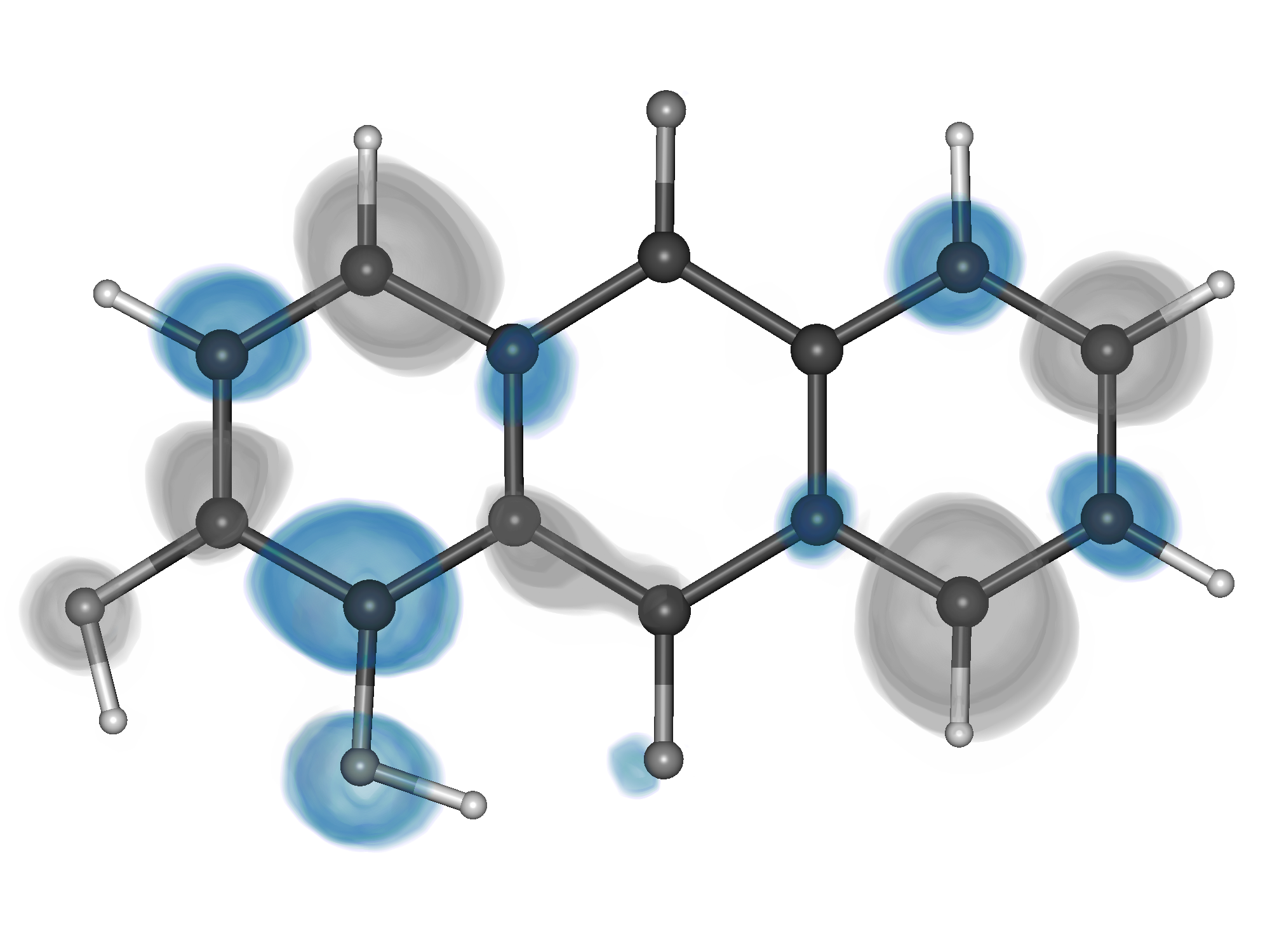}
(c)\includegraphics[width=0.45\textwidth]{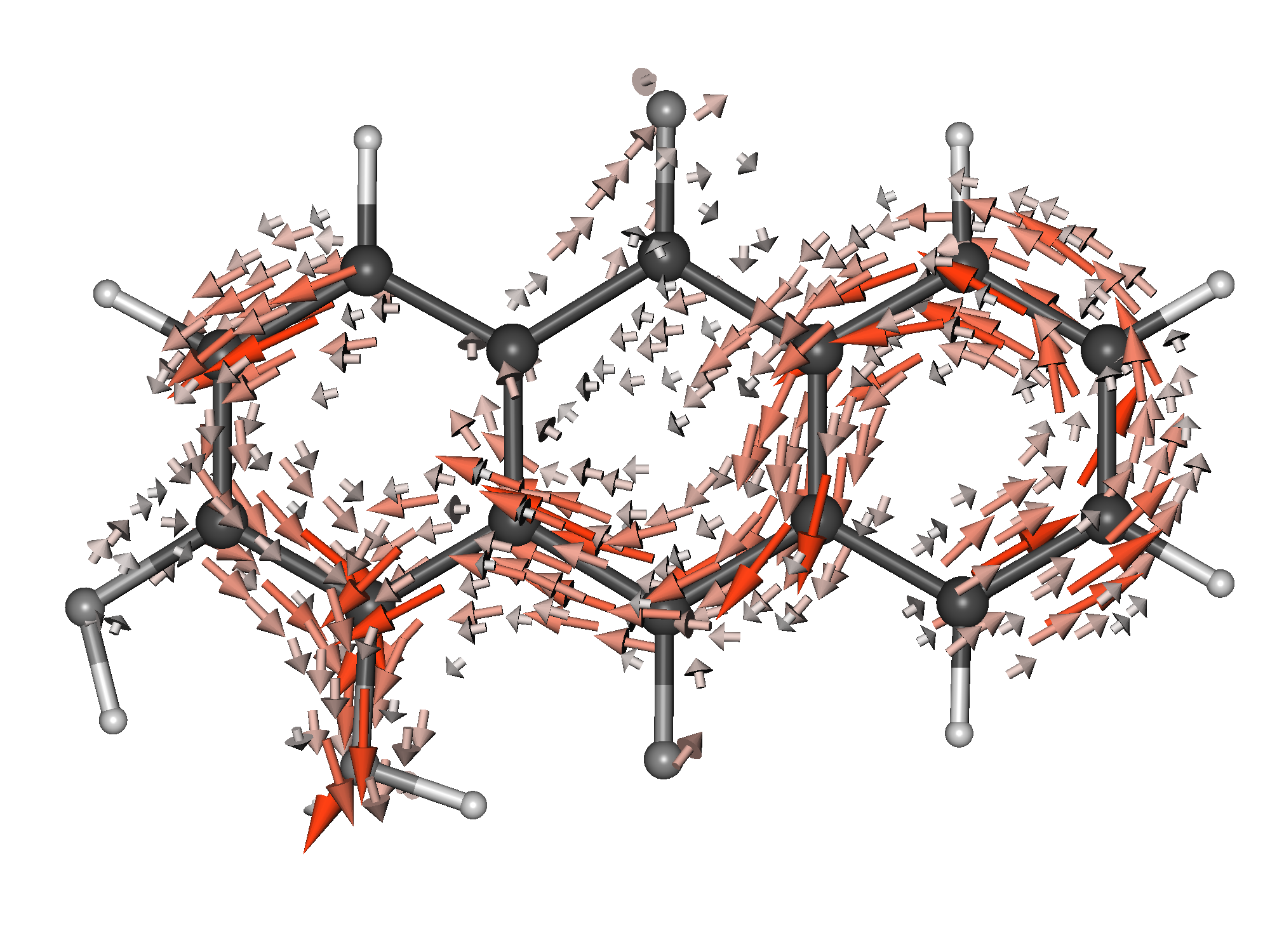}
(d)\includegraphics[width=0.45\textwidth]{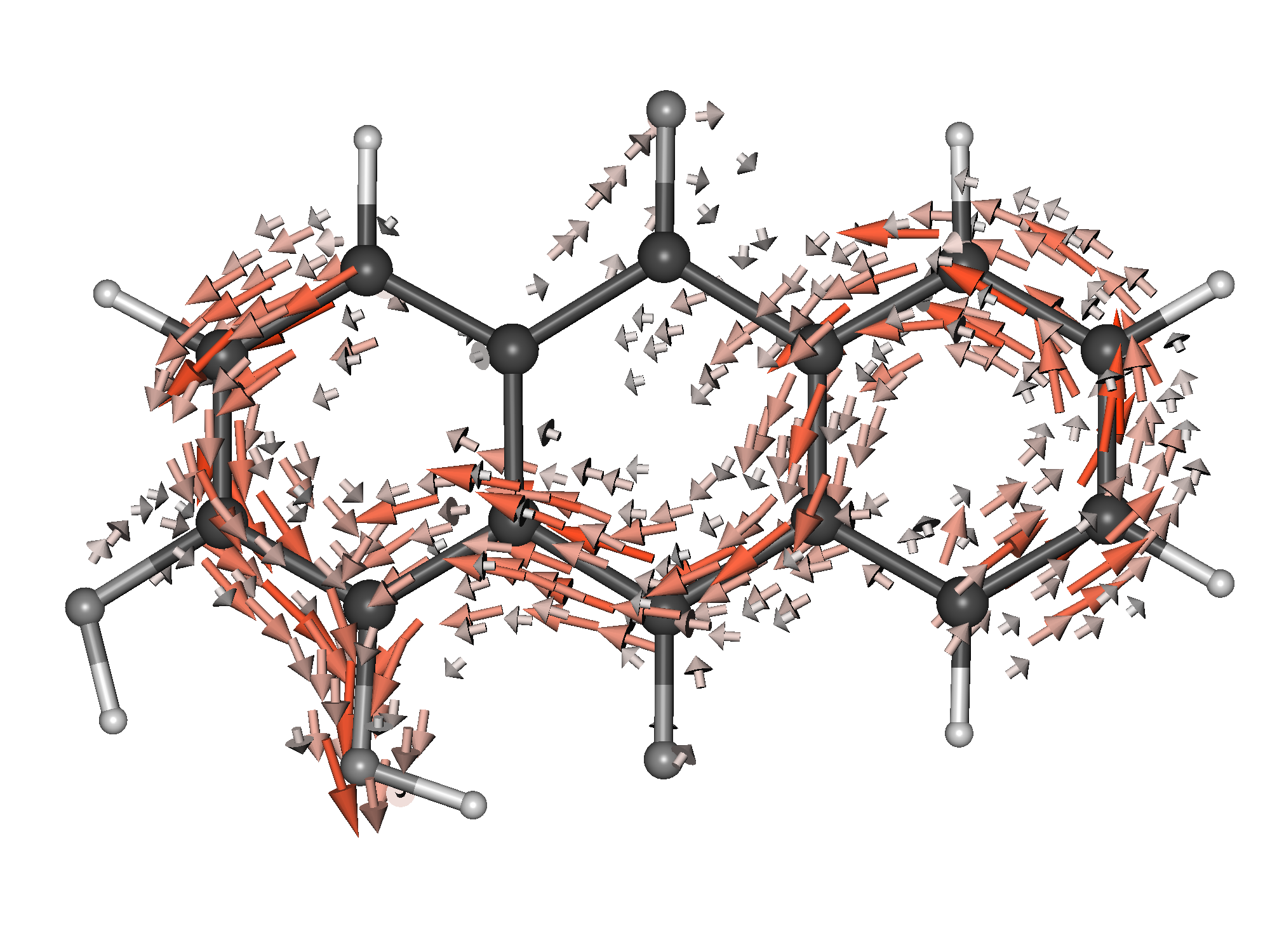}
\caption{Comparison of the electronic flux density $\mathbf{j} \left(\mathbf{r},t\right)$
calculated on the basis of a pseudo-CI eigenfunctions with three different cutoff thresholds, i.e.,
(a) $| C_a^{r,(\lambda)}| > 10^{-3}$, (c) $| C_a^{r,(\lambda)}| > 10^{-2}$, and (d) $| C_a^{r,(\lambda)}| > 10^{-1}$.
The vector arrows are colored according to their magnitude.
In addition, the electron flow $\partial \mathbf{\rho} \left(\mathbf{r},t\right) / \partial t$
calculated with the stringent cutoff threshold is depicted as a contour plot.
The negative and positive isosurfaces are colored gray and blue with a isosurface value of
$\pm3\cdot10^{-4}\,{\rm a_0^{-3}}$.
The time ($t=25.3$\,fs) represents a characteristic turning point during the free decay 
of the electron dynamics for the alizarin dye after broadband laser excitation 
(cf. Fig.~\ref{ali_yld} second time step).
}
\label{ali_flux_dens_comp}
\end{figure*}

To assess the quality of this approximation, the electronic flux densities 
$\mathbf{j} \left(\mathbf{r},t\right)$ reconstructed using the different thresholds are illustrated in Fig.~\ref{ali_flux_dens_comp}(a),(c),(d)
at a characteristic point in time after the laser-pulse excitation (here, $t=25.3$\,fs).
This analysis could be performed during the laser-field at the zeros of the pulse function to avoid simply representing the contribution 
of the electric field to the flux density.
On the other hand, after the pulse, the flux is solely caused by the coherences between the electronic states, simplifying its analysis.
In Fig.~\ref{ali_flux_dens_comp}(b), the corresponding electron flow, $\partial \mathbf{\rho} \left(\mathbf{r},t\right) / \partial t$, is displayed additionally
for the tight threshold, $\left|C_a^{r,(\lambda)}\right| > 10^{-3}$, to facilitate the interpretation of the flux densities.
For all three wave packet expansions, $\mathbf{j} \left(\mathbf{r},t\right)$ shows nearly identical qualitative features.
These include:
(1) an electron flow along the bonds mediated by the $\pi$-system, 
(2) an anti-clockwise flux at the outer right ring, 
and (3) a charge migration from the outer right ring and the top part of the left ring to the lower hydroxyl group.
Due to the cyclic nature of the field-free evolution of the $N$-electron wave packet, the features (2)-(3) exhibits a Rabi-type change of direction
during the dynamics.
Since the qualitative picture is the same even at all levels of analysis, the very important numerical savings at the crudest level of approximation
confer great scalability to the proposed methodology to analyze $N$-electron dynamics in large molecules.
Despite this success, one striking difference can be noticed, i.e., the electron migration 
from the lower hydroxyl group to the neighboring carboxyl group is not fully reproduced 
with the two larger thresholds (cf. Fig. \ref{ali_flux_dens_comp}(c) and (d)).
This corresponds to a through-space charge transfer and is mediated by a large number of small contributions in the pseudo-CI eigenstate basis.
While the major phenomenological characteristics of the correlated electron dynamics are still captured,
some minor mechanistic information is lost by reducing the wave functions to their dominant determinant contributions.

\begin{figure}
\includegraphics[width=0.45\textwidth]{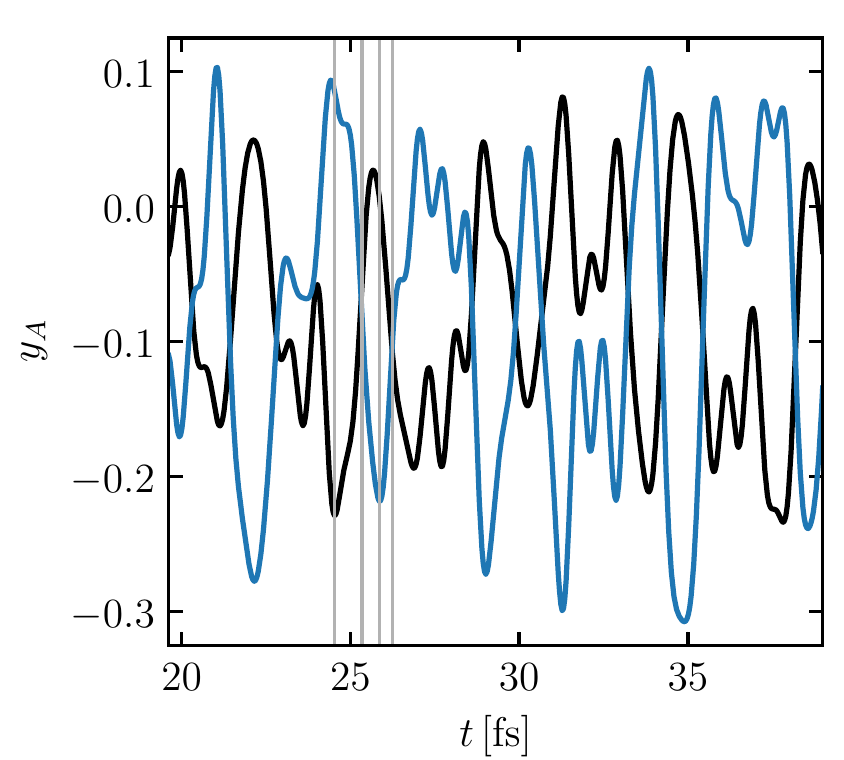}
\caption{Time-evolution of the yield $y_A$ projected 
on the left (black) and right (blue) aromatic ring of the alizarin dye after the 
broadband laser excitation.
The vertical gray lines represent characteristic time points for which the time-dependent 
electronic flux densities $\mathbf{j} \left(\mathbf{r},t\right)$ are evaluated in 
Fig.~\ref{ali_flux_dens}.}
\label{ali_yld}
\end{figure}

\begin{figure*}
(a)\includegraphics[width=0.45\textwidth]{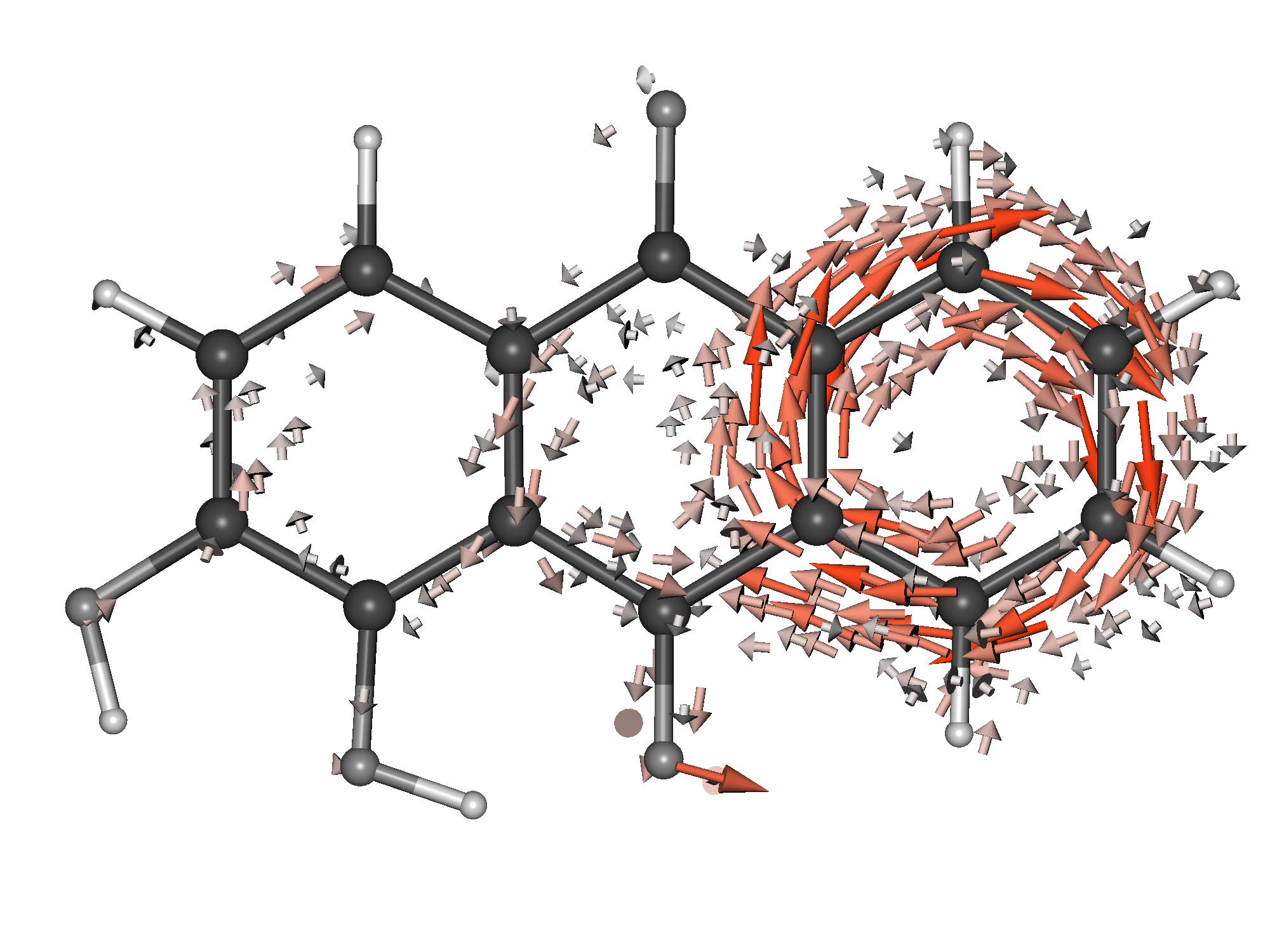}
(b)\includegraphics[width=0.45\textwidth]{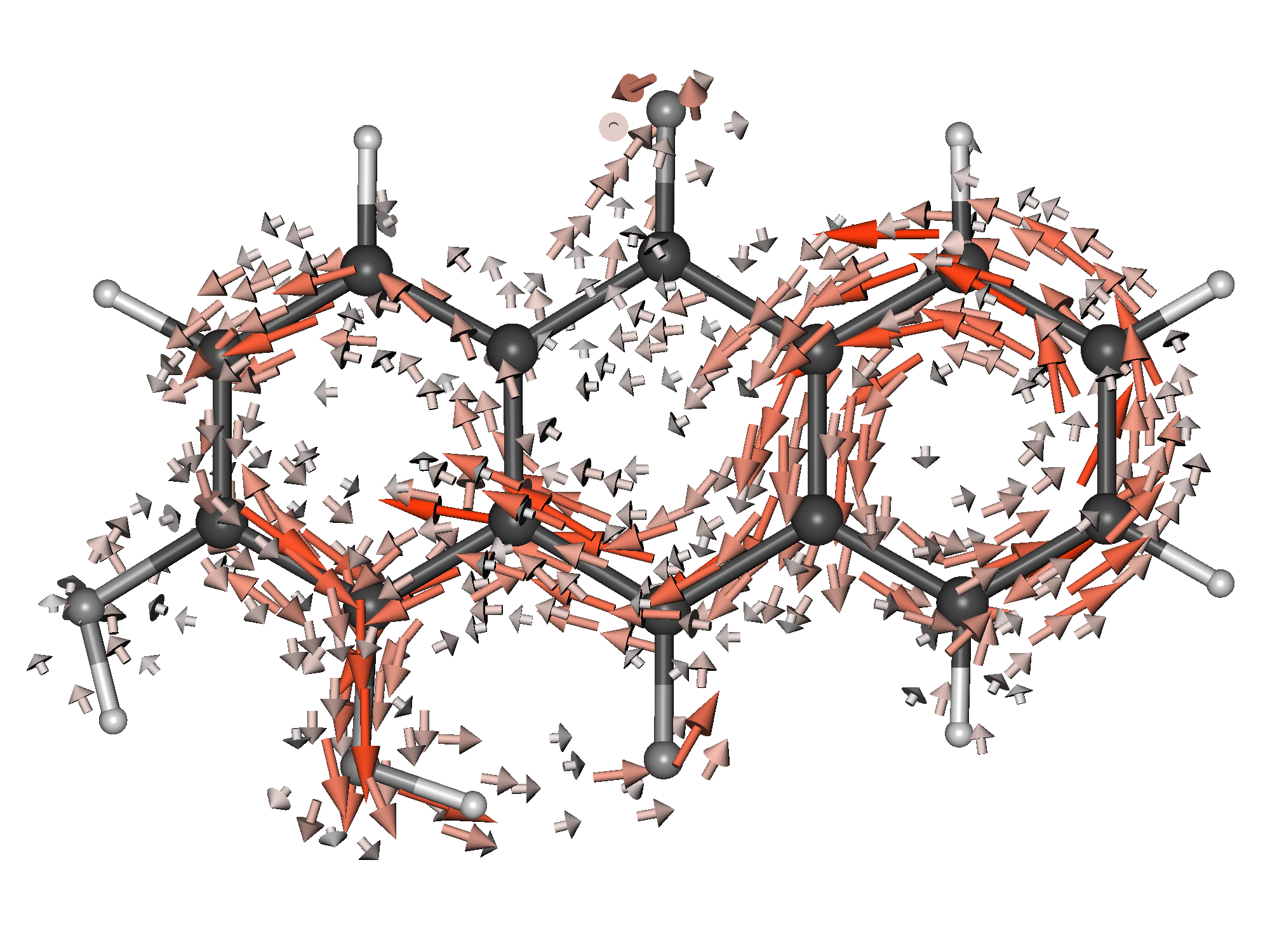}
(c)\includegraphics[width=0.45\textwidth]{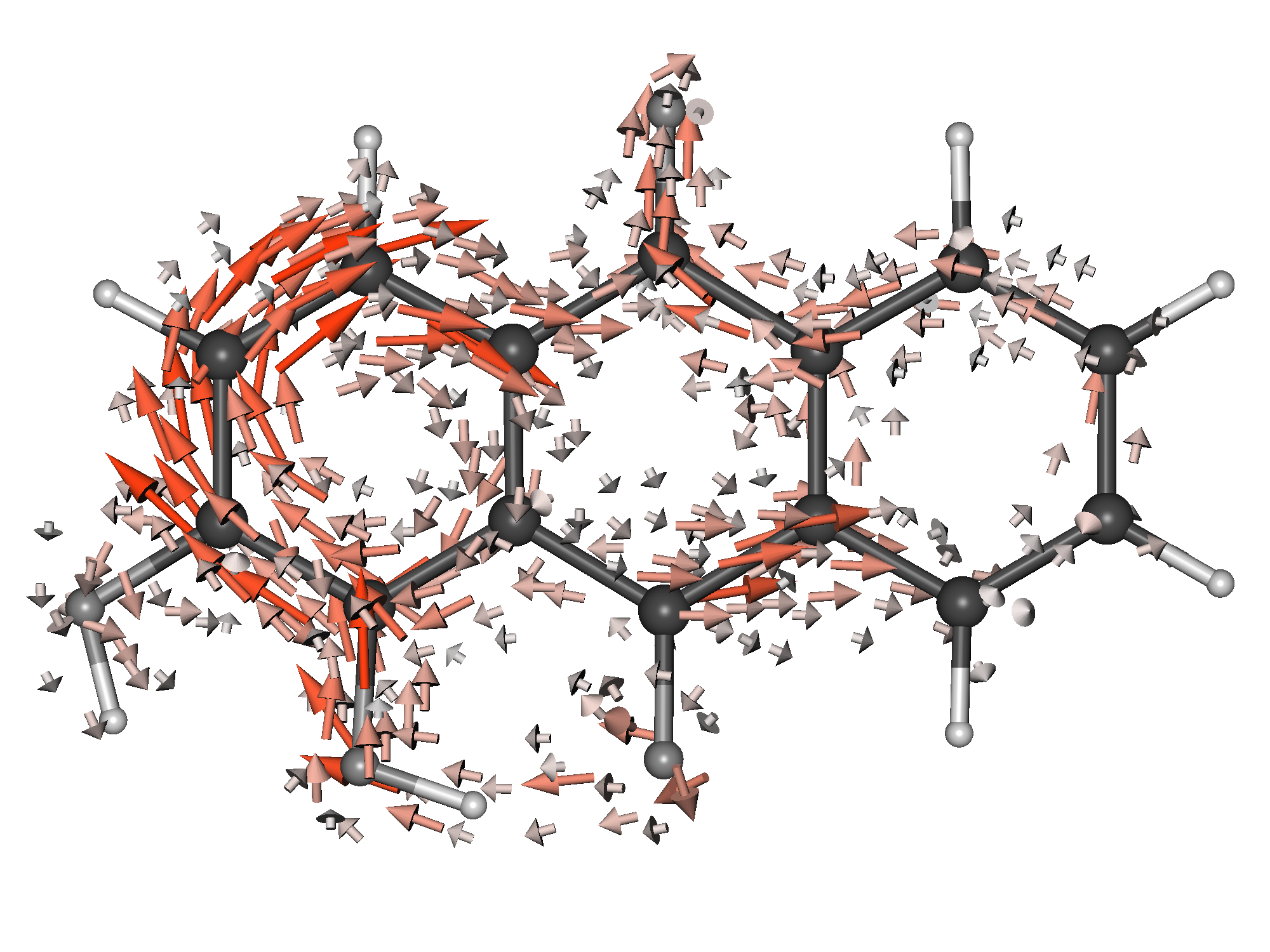}
(d)\includegraphics[width=0.45\textwidth]{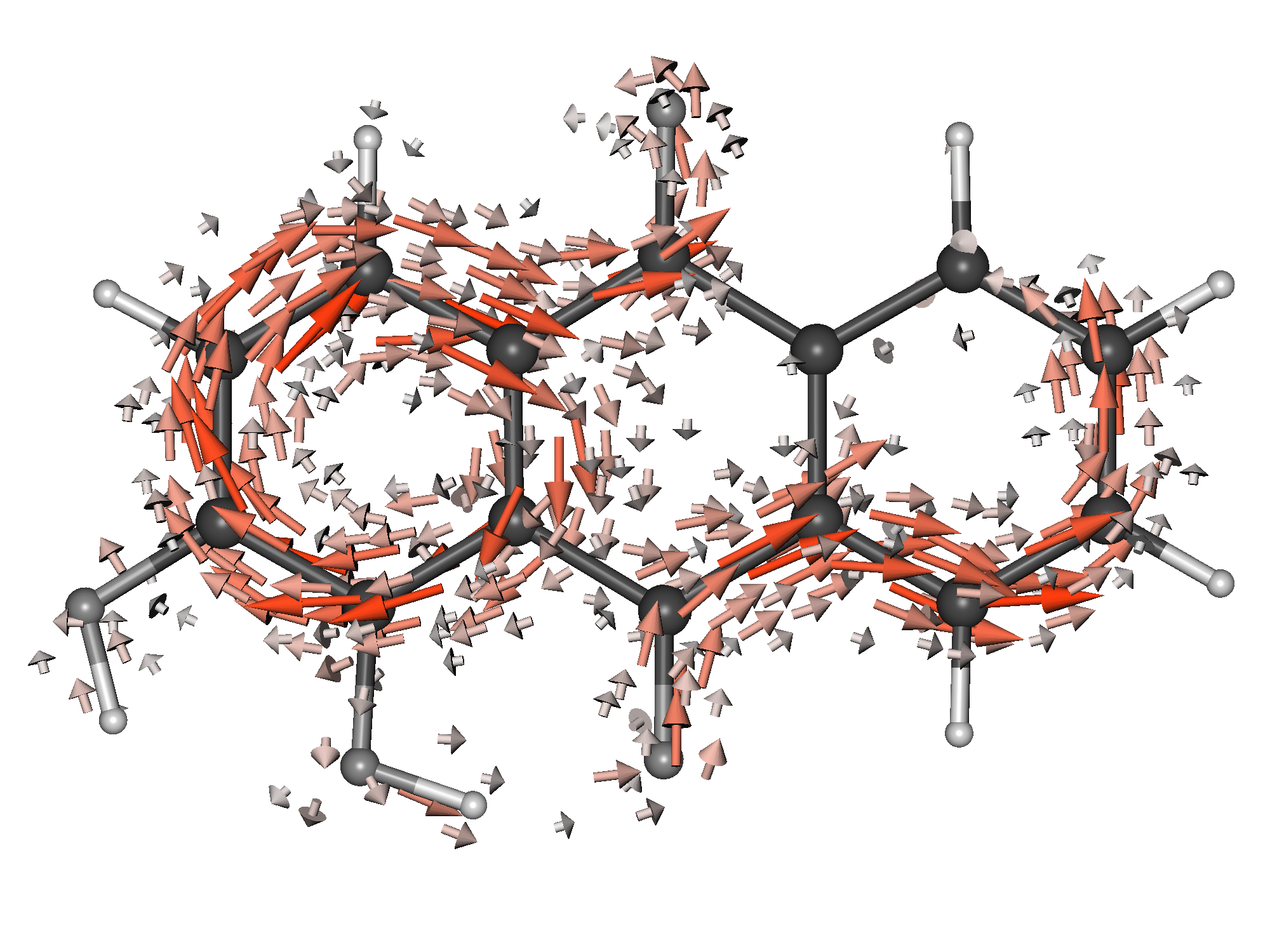}
\caption{Vector plots of the electronic flux density 
$\mathbf{j} \left(\mathbf{r},t\right)$ at representative times for the laser-induced
electron dynamics in alizarin.
The snapshots describe characteristic points of the electron simulation
after the broadband laser excitation and are marked in Fig.~\ref{ali_yld} as gray lines.
The size and color of the vector arrows are scaled according to their magnitude.}
\label{ali_flux_dens}
\end{figure*}

To understand the electron dynamics, we extend its analysis to the time-dependent electronic yield.
It is defined as the difference between the electron density at a given time and the electron density at $t=0\,{\rm fs}$, integrated over a given volume,
\begin{eqnarray}
 y_{\rm A}(t) &=& \int_{\rm A} {\rm d}\mathbf{r} \Big(\rho(\mathbf{r},t) - \rho(\mathbf{r},0) \Big)
\end{eqnarray}
In Fig.~\ref{ali_yld}, the difference densities projected on right and left aromatic ring of alizarin are reported for times after the laser has been switched off.
These reveal intricate synchronous fluctuations of electron density between both rings. The slight asymmetry of the electron redistribution 
correlates with the asymmetric substituents on the two outer rings: the fluctuation amplitudes in the left ring are larger than in the unsubstituted, rightmost aromatic ring.
In Fig.~\ref{ali_flux_dens}, the electronic flux density is plotted for selected characteristic points in order to unveil the mechanism of the charge migration between both rings.
The times associated with these snapshots are marked as vertical gray lines in Fig.~\ref{ali_yld}.
In panel (a), where the electronic yield is largest on the rightmost aromatic ring  ($t=$24.5 fs), the electronic flux density is seen to rotate clockwise,
which will create a temporary local magnetic field.
In the next snapshot (panel (b), $t=$25.3 fs), the  yields in the left and right aromatic rings are equivalent. The electrons are seen to flow laminarly from the right to left aromatic moieties.
Starting from the lowest carbon atom of the right ring, electrons move anticlockwise along the bonds of the right ring, passing by the bottom carbonyl group to bottom hydroxyl group
below the left aromatic ring, and back to the central carbonyl unit via a through space mechanism. This is a strong indication of the presence of a hydrogen bond, which will lead to 
a rapid tautomeric hydrogen transfer. 
At the third time step ($t=$25.9 fs), the electronic yield is maximal in the right ring and the electron flux density rotates clockwise. Contrary to panel (a), the electron flow
is not as strongly localized inside the ring, probably due to the asymmetry of the ring substituents.
In the last snapshot ($t=$26.4 fs), the density migrates back from the leftmost to the rightmost aromatic ring along a different path:
the electrons retain a clockwise orientation in the left ring, mostly flowing from the bottom hydroxyl to the top central carbonyl group. A marginal amount flows from the bottom left 
hydroxy group over the hydrogen bond to the bottom carbonyl group and into rightmost aromatic ring.
For the complete picture of the dynamics after the laser excitation, a short film of the charge migration process is made available online in the Supporting Information.

\section{Conclusions}

In this paper, we have introduced a novel procedure to analyze and visualize many-electron dynamics from a hybrid time-dependent density functional theory (TDDFT)/ configuration interaction singles (CIS) formalism.
The method resorts to a linear-response TDDFT calculation to generate a basis of pseudo-CI eigenvectors and associated energies, which are then used as a basis to describe an $N$-electron dynamics
at the CIS level of theory. The time-dependent CIS wave function retains the simple character of coupled electron-hole pairs, which facilitates its interpretation
in terms of configuration states while keeping the size of the basis relatively small. This renders the hybrid method amenable to large systems -- in fact, any system that can be 
tackled and described accurately using linear-response TDDFT.

From the evolution of a TDDFT/CIS wave packet, we showed how it is possible to rationalize the $N$-electron dynamics of a system in terms of transition moments of various
one-electron operators. These include difference electronic densities, transient electronic flux density maps, and transition dipole moments, which we have implemented in 
a Python toolbox for postprocessing multi-determinantal wave function data. This new module of our open-source project \textsc{ORBKIT} creates a library of transition moments
of user-specified one-electron operators, which are projected on an arbitrary grid. The required information (molecular orbitals, structure, Gaussian atomic basis, pseudo-CI coefficients)
can be directly extracted from the output of various quantum chemistry program packages. These are then used to reconstruct quantities that help understanding the flow of
electrons in molecules out of equilibrium.

We first applied the hybrid TDDFT/CIS scheme to a test system, the charge transfer in lithium hydride, in order to benchmark the quality of the ansatz against standard CIS and Full CI reference simulations.
The results demonstrated that a good choice of the functional improves mostly the energetic description of the charge transfer state, which can be brought close to the Full CI benchmark.
On the other hand, the pseudo-CI basis retains a single excitation character, which is found to be similar to the standard CIS reference. Both ansatz agree semi-quantitatively with the
higher level wave function description, with the discrepancies mostly found in the regions of low density.

In a second example, the application to the broadband excitation of a prototypical chromophore for dye-sensitized solar cells demonstrated the scalability of the method and the versatility of the new toolkit.
In particular, it was found that the main features of the electron flow mechanism can be recovered using a stringent basis pruning strategy, in which each pseudo-CI eigenstate is represented 
using only a few dominant configurations. By doing so, marginal features of the many-electron dynamics involving, e.g., electron flow through hydrogen bonds, may be lost. Because of the favorable 
scaling of the method even at tighter convergence Thresholds, it is expected to be applicable to a large number of medium-sized molecules. This will help to understand electron migration processes
in various fields, such as nano electronics and light-harvesting applications.

\section{Acknowledgment}

The authors gratefully thank the Scientific Computing Services Unit of the 
Zentraleinrichtung f{\"u}r Datenverarbeitung at Freie Universt{\"a}t Berlin for 
allocation of computer time, and Hans-Christian Hege for providing the ZIBAmira 
visualization program. 
The funding of the Deutsche Forschungsgemeinschaft (project TR1109/2-1)
and of the Elsa-Neumann foundation of the Land Berlin are also acknowledged.

\section{Keywords}

Correlated Electron Dynamics, Time-dependent Density Functional Theory, 
Electronic Flux Density, Electron Density, Electronic Current Density

\clearpage




\newpage

\begin{thebibliography}{10}

\bibitem{87:zewail}
M.~Dantus, M.~J. Rosker, and A.~H. Zewail.
\newblock Real-time femtosecond probing of transition states in chemical
  reactions.
\newblock {\em J. Chem. Phys.}, 87:2395, 1987.

\bibitem{88:zewail}
T.~S. Rose, M.~J. Rosker, and A.~H. Zewail.
\newblock Femtosecond real-time observation of wave packet oscillations
  (resonance) in dissociation reactions.
\newblock {\em J. Chem. Phys.}, 88:6672, 1988.

\bibitem{11:SCS:tddftci}
J.~A. Sonk, M.~Caricato, and H.~B. Schlegel.
\newblock Td-ci simulation of the electronic optical response of molecules in
  intense fields: Comparison of {RPA, CIS, CIS(D), and EOM-CCSD}.
\newblock {\em J. Phys. Chem. A}, 115:4678--4690, 2011.

\bibitem{hermann2016alianatase}
G.~Hermann and J.~C. Tremblay.
\newblock Ultrafast photoelectron migration in dye-sensitized solar cells:
  influence of the binding mode and many-body interactions.
\newblock {\em J. Chem. Phys.}, 145:174704, 2016.

\bibitem{klinkusch2016roi}
S.~Klinkusch and J.~C. Tremblay.
\newblock Resolution-of-identity stochastic time-dependent configuration
  interaction for dissipative electron dynamics in strong fields.
\newblock {\em J. Chem. Phys.}, 144(18):184108, 2016.

\bibitem{Pirhadi2016127}
S.~Pirhadi, J.~Sunseri, and D.~R. Koes.
\newblock Open source molecular modeling.
\newblock {\em J. Mol. Graph. Model.}, 69:127, 2016.

\bibitem{Schroedinger1926quantisierung}
E.~Schr\"odinger.
\newblock {Quantisierung als {E}igenwertproblem}.
\newblock {\em Ann. Phys. (Leipzig)}, 81:109, 1926.

\bibitem{slater1929theory}
J.~C. Slater.
\newblock The theory of complex spectra.
\newblock {\em Phys. Rev.}, 34(10):1293, 1929.

\bibitem{condon1930theory}
E.~U. Condon.
\newblock The theory of complex spectra.
\newblock {\em Phys. Rev.}, 36(7):1121, 1930.

\bibitem{slater1931theory}
J.~C. Slater.
\newblock Molecular energy levels and valence bonds.
\newblock {\em Phys. Rev.}, 38:1109, Sep 1931.

\bibitem{orbkit}
G.~Hermann, V.~Pohl, J.~C. Tremblay, B.~Paulus, H.-C. Hege, and A.~Schild.
\newblock \textsc{ORBKIT}: A modular python toolbox for cross-platform
  postprocessing of quantum chemical wavefunction data.
\newblock {\em J. Comput. Chem.}, 37(16):1511, 2016.

\bibitem{Gamess}
Michael~W. Schmidt, Kim~K. Baldridge, Jerry~A. Boatz, Steven~T. Elbert, Mark~S.
  Gordon, Jan~H. Jensen, Shiro Koseki, Nikita Matsunaga, Kiet~A. Nguyen, Shujun
  Su, Theresa~L. Windus, Michel Dupuis, and John~A. Montgomery.
\newblock General atomic and molecular electronic structure system.
\newblock {\em J. Comput. Chem.}, 14(11):1347, 1993.

\bibitem{TURBOMOLE}
{TURBOMOLE V6.5}, a development of {University of Karlsruhe} and
  {Forschungszentrum Karlsruhe GmbH}, 1989-2007, {TURBOMOLE GmbH}, since 2007,
  2013.
\newblock available via {http://www.turbomole.com}.

\bibitem{ci_orbkit_wf}
V.~{Pohl}, G.~{Hermann}, and J.~C. {Tremblay}.
\newblock {An Open-Source Framework for Analyzing $N$-Electron Dynamics: I.
  Multi-Determinantal Wave Functions}.
\newblock {\em ArXiv e-prints arXiv:1701.06885 [physics.chem-ph]}, January 2017.

\bibitem{wavepacket}
Ulf~Lorenz B.~Schmidt.
\newblock {WavePacket 5.2}, : A matlab program package for quantum-mechanical
  wavepacket and density propagation and time-dependent spectroscopy., 2016.
\newblock Available via {http://sourceforge.net/p/wavepacket/matlab}.

\bibitem{tremblay2008time}
J.~C. Tremblay, T.~Klamroth, and P.~Saalfrank.
\newblock Time-dependent configuration-interaction calculations of laser-driven
  dynamics in presence of dissipation.
\newblock {\em J. Chem. Phys.}, 129(8):084302, 2008.

\bibitem{08:TS:gloct}
J.~C. Tremblay and P.~Saalfrank.
\newblock Guided locally-optimal control of quantum dynamics in dissipative
  environments.
\newblock {\em Phys. Rev. A}, 78:063408:1--9, 2008.

\bibitem{psi4}
J.~M. Turney, A.~C. Simmonett, R.~M. Parrish, E.~G. Hohenstein, F.~A.
  Evangelista, J.~T. Fermann, B.~J. Mintz, L.~A. Burns, J.~J. Wilke, M.~L.
  Abrams, N.~J. Russ, M.~L. Leininger, C.~L. Janssen, E.~T. Seidl, W.~D. Allen,
  H.~F. Schaefer, R.~A. King, E.~F. Valeev, C.~D. Sherrill, and T.~D. Crawford.
\newblock Psi4: an open-source ab initio electronic structure program.
\newblock {\em WIREs Comput. Mol. Sci.}, 2(4):556, 2012.

\bibitem{foresman1992toward}
J.~B. Foresman, M.~Head-Gordon, J.~A. Pople, and M.~J. Frisch.
\newblock Toward a systematic molecular orbital theory for excited states.
\newblock {\em J. Phys. Chem.}, 96(1):135, 1992.

\bibitem{dreuw2003long}
A.~Dreuw, J.~L. Weisman, and M.~Head-Gordon.
\newblock Long-range charge-transfer excited states in time-dependent density
  functional theory require non-local exchange.
\newblock {\em J. Chem. Phys.}, 119(6):2943, 2003.

\bibitem{dreuw2005single}
A.~Dreuw and M.~Head-Gordon.
\newblock Single-reference ab initio methods for the calculation of excited
  states of large molecules.
\newblock {\em Chem. Rev.}, 105(11):4009, 2005.

\bibitem{gross1990time}
E.~K.~U. Gross and W.~Kohn.
\newblock Time-dependent density-functional theory.
\newblock {\em Adv. Quantum Chem.}, 21:255, 1990.

\bibitem{casida1998molecular}
M.~E. Casida, C.~Jamorski, K.~C. Casida, and D.~R. Salahub.
\newblock Molecular excitation energies to high-lying bound states from
  time-dependent density-functional response theory: Characterization and
  correction of the time-dependent local density approximation ionization
  threshold.
\newblock {\em J. Chem. Phys.}, 108(11):4439, 1998.

\bibitem{iikura2001long}
H.~Iikura, T.~Tsuneda, T.~Yanai, and K.~Hirao.
\newblock A long-range correction scheme for generalized-gradient-approximation
  exchange functionals.
\newblock {\em J. Chem. Phys.}, 115(8):3540, 2001.

\bibitem{heyd2003hybrid}
J.~Heyd, G.~E. Scuseria, and M.~Ernzerhof.
\newblock Hybrid functionals based on a screened coulomb potential.
\newblock {\em J. Chem. Phys.}, 118(18):8207, 2003.

\bibitem{tawada2004long}
Y.~Tawada, T.~Tsuneda, S.~Yanagisawa, T.~Yanai, and K.~Hirao.
\newblock A long-range-corrected time-dependent density functional theory.
\newblock {\em J. Chem. Phys.}, 120(18):8425, 2004.

\bibitem{rohrdanz2008simultaneous}
M.~A. Rohrdanz and J.~M. Herbert.
\newblock Simultaneous benchmarking of ground-and excited-state properties with
  long-range-corrected density functional theory.
\newblock {\em J. Chem. Phys.}, 129(3):034107, 2008.

\bibitem{rohrdanz2009long}
M.~A. Rohrdanz, K.~M. Martins, and J.~M. Herbert.
\newblock A long-range-corrected density functional that performs well for both
  ground-state properties and time-dependent density functional theory
  excitation energies, including charge-transfer excited states.
\newblock {\em J. Chem. Phys.}, 130(5):054112, 2009.

\bibitem{jacquemin2009extensive}
D.~Jacquemin, V.~Wathelet, E.~A. Perpete, and C.~Adamo.
\newblock Extensive {TD-DFT} benchmark: singlet-excited states of organic
  molecules.
\newblock {\em J. Chem. Theory Comput.}, 5(9):2420, 2009.

\bibitem{becke1988density}
A.~D. Becke.
\newblock Density-functional exchange-energy approximation with correct
  asymptotic behavior.
\newblock {\em Phys. Rev. A}, 38(6):3098, 1988.

\bibitem{schafer1992fully}
A.~Sch{\"a}fer, H.~Horn, and R.~Ahlrichs.
\newblock Fully optimized contracted gaussian basis sets for atoms {Li} to
  {Kr}.
\newblock {\em J. Chem. Phys.}, 97(4):2571, 1992.

\bibitem{weigend2005balanced}
F.~Weigend and R.~Ahlrichs.
\newblock Balanced basis sets of split valence, triple zeta valence and
  quadruple zeta valence quality for {H} to {Rn}: design and assessment of
  accuracy.
\newblock {\em Phys. Chem. Chem. Phys.}, 7(18):3297, 2005.

\bibitem{duncan2005electronic}
W.~R. Duncan and O.~V. Prezhdo.
\newblock Electronic structure and spectra of catechol and alizarin in the gas
  phase and attached to titanium.
\newblock {\em J. Phys. Chem. B}, 109(1):365, 2005.

\bibitem{sanchez2010real}
R.~S{\'a}nchez-de Armas, J.~Oviedo~L{\'o}pez, M.~A.~San-Miguel, J.~F. Sanz,
  P.~Ordej{\'o}n, and M.~Pruneda.
\newblock Real-time {TD-DFT} simulations in dye sensitized solar cells: the
  electronic absorption spectrum of alizarin supported on {TiO$_2$}
  nanoclusters.
\newblock {\em J. Chem. Theory Comput.}, 6(9):2856, 2010.

\bibitem{Gomez_2015}
T.~Gomez, G.~Hermann, X.~Zarate, J.~P{\'e}rez-Torres, and J.~C. Tremblay.
\newblock Imaging the ultrafast photoelectron transfer process in
  alizarin-{TiO$_2$}.
\newblock {\em Molecules}, 20(8):13830, Jul 2015.

\bibitem{huber2002real}
R.~Huber, J.-E. Moser, M.~Gr{\"a}tzel, and J.~Wachtveitl.
\newblock Real-time observation of photoinduced adiabatic electron transfer in
  strongly coupled dye/semiconductor colloidal systems with a 6 fs time
  constant.
\newblock {\em J. Phys. Chem. B}, 106(25):6494, 2002.

\bibitem{dworak2009ultrafast}
L.~Dworak, V.~V. Matylitsky, and J.~Wachtveitl.
\newblock Ultrafast photoinduced processes in alizarin-sensitized metal oxide
  mesoporous films.
\newblock {\em ChemPhysChem}, 10(2):384, 2009.

\bibitem{tremblay2004using}
J.~C. Tremblay and T.~Carrington~Jr.
\newblock Using preconditioned adaptive step size {Runge-Kutta} methods for
  solving the time-dependent {Schr{\"o}dinger equation}.
\newblock {\em J. Chem. Phys.}, 121(23):11535, 2004.

\bibitem{tremblay2011dissipative}
J.~C. Tremblay, S.~Klinkusch, T.~Klamroth, and P.~Saalfrank.
\newblock Dissipative many-electron dynamics of ionizing systems.
\newblock {\em J. Chem. Phys.}, 134(4):044311, 2011.

\end{thebibliography}
\end{document}